\documentclass[prl,aps,twocolumn,twoside,superscriptaddress]{revtex4}

\usepackage{amssymb}
\usepackage{amsmath}
\usepackage{amscd}
\usepackage{latexsym}  
\usepackage{epsfig}
\usepackage{graphicx}
\usepackage{bbm}

\newtheorem{defi}{Definition}

\newtheorem{exempel}[defi]{Example}

		\def\CC{{\cal C}}
		
	\def\CH{{\cal H}}

	\def\CN{{\cal N}}	
	\def\CQ{{\cal Q}}	
\def\CS{{\cal S}}	\def\CT{{\cal T}}

\def\beq{\begin {equation}}
\def\eeq{\end {equation}}

\newcommand{\tr}{{\operatorname{Tr}}}

\newcommand{\bra}[1]{{\langle{#1}|}}
\newcommand{\ket}[1]{{|{#1}\rangle}}
\newcommand{\ketbra}[1]{{\ket{#1}\!\bra{#1}}}

\newcommand{\1}{{\openone}}

\def\bar{\overline}
\newlength{\blank}
\newlength{\equalsign}
\begin{document}

\title{Relating quantum privacy and quantum coherence: an operational approach}
\date{July 8, 2003}
\author{I. Devetak}
\affiliation{IBM T.~J.~Watson Research Center, PO Box 218, Yorktown Heights, NY 10598, USA}
\author{A. Winter}
\affiliation{Department of Computer Science, University of Bristol, Bristol BS8 1UB, U.K.}

\maketitle

{\small\it {Abstract} ---
     We describe how to achieve optimal entanglement generation and one-way 
     entanglement distillation rates by coherent implementation of a class of 
     secret key generation and secret key distillation protocols, respectively.}

\par\medskip

Quantum information theory may be understood in terms of inter-conversion
between various resources \cite{CR}. 
One very desirable resource is 
a maximally entangled state shared between two parties (Alice and Bob).
It allows them to perform various tasks such as teleportation, 
super-dense coding, etc.~\cite{tasks}.
Another is 
a pair of maximally correlated classical bit-strings 
shared by the two parties, reliably secret from any third party (Eve). 
It has important cryptographic applications since it can be used as a 
\emph{secret key} (or ``one time pad'') allowing Alice and Bob private 
communication over a public channel.
Although used for different purposes, the two resources are intimately 
related through the property of ``exclusiveness''
--- in the one case toward the total outside world, in the other toward 
the eavesdropper Eve. This connection has been confirmed by a growing literature of 
useful analogies \cite{collins:popescu, SW:private} and operational
equivalences \cite{shor:preskill,acin:key}.
In this Letter we further exploit this connection
to relate several private and quantum protocols, culminating in the
demonstration of the ``hashing inequality'' for one-way entanglement distillation
\cite{HHH,BDSW}. We consider four distinct resource conversion scenarios, 
starting from  
a noisy quantum channel/entanglement and ending up with
near-perfect secret key/entanglement.
In other words we are converting a noisy dynamic/static quantum resource 
into a noiseless static private/quantum resource.
\par\medskip\noindent
1. static $\rightarrow$ private = \emph{Secret key distillation.} The task is to
convert $n$ copies of the bipartite state $\rho^{A B}$
shared between Alice and Bob into $n R$ bits of 
secret key using 1-LOPC (local quantum operations and forward public communication). 
The quantity $R$ is referred to as the \emph{rate} of the protocol.
To understand what is meant by a secret key, consider a purification 
${\psi}^{ABE} =  \ket{\psi}\bra{\psi}^{ABE}$ of $\rho^{AB}$ and allow for the 
worst case
scenario in which the eavesdropper Eve is given the purifying 
system $E$. Defining $\bar{\Phi} = \frac{1}{2} (\ket{00} \bra{00} + 
\ket{11} \bra{11})$, a pair of maximally correlated bits written formally as a 
density operator, the desired shared secret key is represented as
${(\bar{\Phi}^{AB})}^{\otimes n R} \otimes \theta^E$; the classical
key shared between Alice and Bob is decoupled from Eve's state $\theta$.
\par\medskip\noindent
2. static $\rightarrow$ quantum  = \emph{Entanglement distillation.} Here, 
by  1-LOCC (local quantum operations and forward classical communication)
$(\rho^{AB})^{\otimes n}$ is to be converted into $nR$ bits of entanglement
$\Phi_+^{\otimes n R}$,  where 
$\ket{{\Phi_+}} = \frac{1}{\sqrt{2}} (\ket{00} + \ket{11})$.
Since the final state is pure,  the decoupling from $E$ is \emph{implicit}.
Including $E$ in the description, the effect of the protocol
may be written as  
${(\psi^{ABE})}^{\otimes n} \Longrightarrow 
    {(\Phi^{AB}_+)}^{\otimes n R} \otimes \theta^E$,
which is just a coherent version of its private counterpart above!  
\par\medskip\noindent
3. dynamic $\rightarrow$ private = \emph{Secret key generation.}
Here the starting point is a noisy quantum channel $\CN$, a cptp map
taking density operators in the Hilbert space of Alice's system $A'$ to 
that of Bob's $B$.
Applying $\CN$ is equivalent to an isometric
mapping onto a larger Hilbert space corresponding to $BE$, which includes
the \emph{environment} $E$ assumed at Eve's disposal. 
The channel may be characterized by its effect on some bipartite pure state
$\ket{\psi'}^{A A'}$, living entirely on Alice's side, which now
becomes the mixed state $\rho^{AB} = (\1^A \otimes \CN) \psi'$
shared between her and Bob. Note that the purifying system for 
$\rho^{AB}$ is precisely $E$, so that we again have the tripartite state
${\psi}^{ABE}$. The task is, by Alice choosing an appropriate
input to $\CN^{\otimes n}$ and Bob performing a decoding operation, to
generate $nR$ bits of secret key, decoupled from $E$ as in the first scenario.
\par\medskip\noindent
4. dynamic $\rightarrow$ quantum = \emph{Entanglement generation.} The channel $\CN$
is now used to generate $nR$ bits of entanglement. Again one
may explicitly include $E$ to stress the resemblance to the private scenario. 
\par\medskip
We shall show how a particular asymptotic rate $R$ defined in terms
of the state ${\psi}^{ABE}$ may be \emph{achieved} for the four scenarios,
by which we mean that for sufficiently large $n$ there exists a protocol whose
output approximates the desired state arbitrarily closely (in trace distance).  
First we give a simple dimension counting argument for 
secret key generation, and show how to augment it to work
for secret key distillation. Then we modify the two protocols to
make them ``coherent'',
yielding the entanglement generation and distillation protocols, respectively.
Finally, we point to a more direct connection between the two coherent
protocols and discuss the implications of our results for
finding \emph{optimal} rates. A rigorous treatment of
the dynamic and static scenarios via random code selection may be found in
\cite{devetak} and \cite{devetak:winter}, respectively, 
building on results from classical information theory \cite{AC:1}.
\par 
The state ${\psi}^{ABE}$, related to our coherent protocols, 
may be written in the Schmidt representation with respect to the $A|BE$ partition
as
$$
\ket{\psi}^{ABE} = \sum_x \sqrt{P(x)} \ket{x}^A \otimes \ket{\phi_x}^{B E}.
$$
In the dynamic scenario it comes about by sending some 
$\ket{\psi}^{A A'} = \sum_x \sqrt{P(x)} \ket{x}^A \otimes \ket{\phi'_x}^{A'}$
through the channel.
Relevant to our private protocols is the ``decohered'' state
\beq
\bar{\psi}^{ABE} = \sum_x P(x) \ket{x}\bra{x}^A \otimes \phi^{B E}_x,
\label{eq:cqq}
\eeq
obtainable from $\ket{\psi}^{ABE}$ by measurement in the $\{ \ket{x} \}$ basis.
It represents the state of
a \emph{cqq system} $XBE$ \cite{CR}, $X$ being a random variable
with  probability distribution $P(x)$.
In the dynamic scenario it arises from sending the ensemble 
$\{ P(x), \ket{\phi'_x}^{A'} \}$ through the channel.
\par Define $\omega = \bar{\psi}^B = {\psi}^B $ and 
$\sigma = \bar{\psi}^E = {\psi}^E  $.
For a quantum state $\rho$ we denote the von Neumann entropy 
$H(\rho)=-\tr\rho\log\rho$, and the Shannon
entropy of a random variable $X$, $H(X)=-\sum_x P(x)\log P(x)$.
If the state is the reduced state of a multi--party state, like
the $\bar{\psi}^{ABE}$ above, we write $H(A)=H(\bar{\psi}^A)$, etc. In the particular
case of (\ref{eq:cqq}), obviously $H(\bar{\psi}^A)=H(X)$.
For a general bipartite state on $AB$ define the
conditional entropy $H(A|B) = H(AB) - H(B)$ and 
quantum mutual information
$$I(A;B)=H(A)+H(B)-H(AB).$$
The latter for the cqq state in (\ref{eq:cqq}) is easily checked to be equal to
the \emph{Holevo information}
$H(\omega) - \sum_x P(x)H({\phi}^B_x)$ \cite{Holevo}
for which we shall use the notation $I(X;B)$. 
As a rule, information theoretical quantities involving $X$ are
implicitly referring to $\bar{\psi}^{ABE}$, and those involving
$A$ to  ${\psi}^{ABE}$.
For instance, the \emph{coherent information}
$I_c(A \rangle B) = -H(A|B)$ \cite{coherent}
refers to ${\psi}^{ABE}$, but may be written as
$H(B) - H(E) = I(X;B) - I(X;E)$ when referring to $\bar{\psi}^{ABE}$.
The protocols we are about to describe will all achieve the rate 
$R = I_c(A \rangle B)$. In the entanglement distillation scenario
this is known as the ``hashing inequality'' \cite{HHH, BDSW}.
\paragraph{Typicality.} Let us review the properties of typical sequences 
and subspaces. For the random variable $X$,
any $\epsilon, \delta > 0$ and  sufficiently large $n$ 
there exists a \emph{typical set} $T^n_{X, \delta}$ consisting of sequences 
$x^n = x_1 x_2 \dots x_n$ of length $n$ such that 
$$2^{n [H(X) - \delta]} \leq |T^n_{X,\delta}|  \leq 2^{n [H(X) + \delta]},$$
and $\Pr\{X^n \in T^n_{X, \delta} \} \geq 1 - \epsilon$.   
Typical sequences are those in which the fraction of a given
letter $x$ is approximated by its probability $P(x)$, and the law of large
numbers guarantees that such sequences will occur with high probability.
In less formal notation, large $n$ will be implicit, the typical
set denoted by $T_X$, $H(X) \pm \delta$ written as $H(X)^{\pm}$ and
$1 - \epsilon$ as $\approx 1$.
\par
The quantum analogue of the typical set is the \emph{typical subspace}
\cite{typical} $\CT_B$ of the Hilbert space $\CH_B^{\otimes n}$,
defined for the quantum system $B$ in the state $\omega$. 
It satisfies
$$2^{n H(B)^-} \leq \dim \CT_{B} \leq 2^{n H(B)^+},$$
and $\omega^{\otimes n}$ is approximately
supported on $\CT_B$ in the sense that
$\Pi_{B} \omega^{\otimes n} \Pi_{B} \approx \omega^{\otimes n}$ (in trace
distance) with 
$\Pi_{B}$ the projector onto $\CT_{B}$.
\par
For a cq system $XB$ 
and a particular sequence $x^n \in T_{X}$ there exists a 
\emph{conditionally typical subspace} $\CT_{B|X}(x^n)$
on which $\phi^B_{{x^n}} = \bigotimes_i \phi^B_{x_i}$ is approximately
supported, such that 
$$2^{n H(B|X)^-} \leq \dim \CT_{B|X}(x^n)
 \leq 2^{n H(B|X)^+}.$$
Another important fact is that the $\CT_{B|X}(x^n)$ can be thought of as 
being approximately contained
in $\CT_{B}$  in the sense that $\Pi_{B} \Pi_{B|X}(x^n) \Pi_{B}$ 
enjoys the same asymptotic properties as $\Pi_{B|X}(x^n)$.
 \paragraph {HSW codes.} An HSW \cite{Holevo:coding}  code  $\CC$ 
associated with the Alice-Bob cq system $XB$
is a subset of $T_X$ such that the states $(\psi^B_{x^n})_{x^n \in \CC}$ 
can be distinguished with probability $\approx 1$.
Intuitively, the $\psi^B_{x^n}$, supported on the respective $\CT_{B|X}(x^n)$ of 
dimension at most $2^{n H(B|X)^+}$, can be ``packed'' into $\CT_B$ of dimension
at least $2^{n H(B)^-}$ with negligible overlap if we take 
$|\CC| = 2^{n I(X;B)^-}$ (Fig. 1). The HSW theorem \cite{Holevo:coding} 
confirms this geometric picture.
\begin{figure}[ht]
    \label{fig:babushka}
    \includegraphics[width=7.5cm]{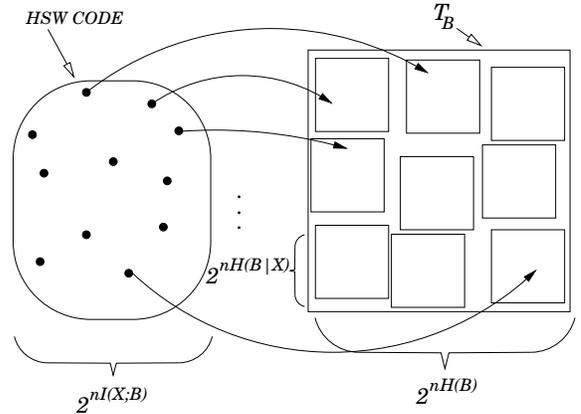}
    \caption{An HSW code. The arrows represent the map $x^n \mapsto \phi^B_{x^n}$. }
\end{figure}
\vspace{-0.25cm}
 \paragraph{Privacy amplification sets.}
Privacy amplification (PA) sets  are, in a sense, dual to HSW codes.
A privacy amplification set $\CS$, 
associated with the Alice-Eve cq system $XE$,
is a subset of $T_X$ of size $S$ such that 
\beq
\frac{1}{S} \sum_{x^n \in \CS} \phi^E_{x^n} \approx {\sigma}^{\otimes n}.
\label{zadovolji}
\eeq
Evidently $\CS = T_X$ is a valid PA set, but we are interested in making
$S$ as small as possible. 
Since $\sigma^{\otimes n}$ and $\phi^E_{x^n}$ are ``almost'' uniformly
supported on $\CT_{E}$ and $\CT_{E|X}(x^n)$, respectively,
satisfying (\ref{zadovolji}) amounts to ``covering'' $\CT_{E}$, 
of dimension at most  $2^{n H(E)^+}$ with the $\CT_{E|X}(x^n)$, $x^n \in \CS$, 
each of which has dimension at least $2^{n H(E|X)^-}$ (Fig. 2). 
It can be shown (using techniques from \cite{AW}) that the quotient of these
dimensions, $S = 2^{n I(X;E)^+}$, indeed suffices.
\begin{figure}[ht]
    \label{fig:babushka2}
    \includegraphics[width=7.5cm]{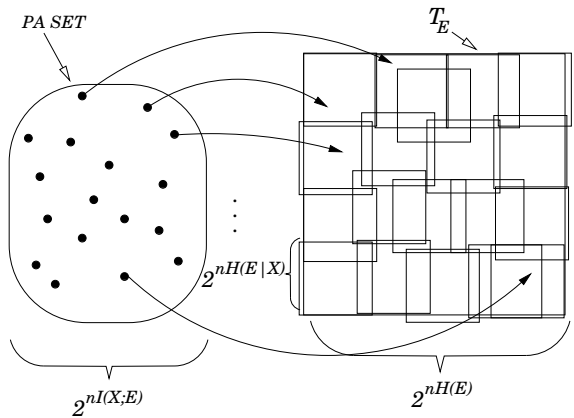}
    \caption{A privacy amplification set. The arrows represent the map 
             $x^n \mapsto \phi^E_{x^n}$. }
\end{figure}
\vspace{-0.25cm}
 \paragraph{Key generation (3.).} We have seen that
randomizing over the members of the PA set leaves Eve with a state
essentially independent of the choice of $\CS$. 
The key generation strategy is now clear.
Construct an HSW code $ (u^{ms})_{m,s}$, $m = 1, \dots M$, $M = 2^{nR^-}$,
$s = 1, \dots S$,  of size $MS = 2^{n I(X;B)^-}$ such that each
$\CS_m = (u^{ms})_s$ is a PA set. The key generation code is such a partitioned
HSW code (Fig. 3).
Alice sends the ensemble $\{ \frac{1}{M}, \frac{1}{S} \sum_s {\phi'}_{u^{ms}}\}$
through the channel, resulting in the cq state
$$
\bar{\Psi}^{A B E} =  \frac{1}{M}
\sum_m \ket{m}\bra{m}^{A} \otimes \frac{1}{S} \sum_s \phi_{u^{ms}}^{BE}.
$$
Bob measures his system to find out $m$
and Eve is left with the state $\theta^E \approx \sigma^{\otimes n}$ 
independent of the index $m$.  
 \paragraph{Key distillation (1.).}
The geometric idea (Fig. 3) is to  cover the space $T_X$ of
dimension at most $2^{n H(X)^+}$ with $L = 2^{n H(X|B)^+}$
key generation codes  $\CC_l = (u^{lms})_{m,s}$ 
of size at least $2^{n I(X;B)^-}$.
Alice converts $\psi^{ABE}$ into $\bar{\psi}^{ABE}$ for all $n$ copies
by measurement in the $\{ \ket{x} \}$ basis. 
The measurement outcome $x^n$ with probability $\approx 1$ lies
in $T_X$. Due to the covering, $x^n$ lies in some key generation 
code labeled by $l$. Alice sends the 
``which key generation code'' information $l$ (this requires $n I(X|B)^+$ bits),
leaves the overall system in the state $\bar{\Psi}^{A B E}$ \cite{remarq}
and Bob simply proceeds as in key generation.
The extra classical communication thus compensates 
for the initial resource being static rather than dynamic.
\par
We now show how the two private protocols can be made coherent.
\begin{figure}[ht]
    \label{fig:babushka3}
    \includegraphics[width=8.5cm]{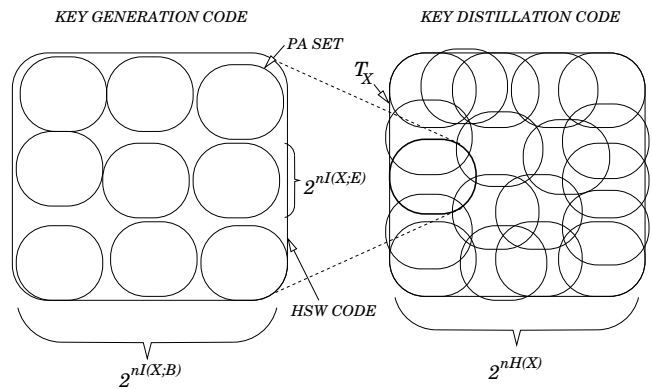}
    \caption{A secret key generation code (left) viewed as a building 
             block for a key distillation code (right).}
\end{figure}
 \paragraph{Entanglement generation (4.).} For entanglement generation
Alice prepares
$
\frac{1}{\sqrt{M}} \sum_m \ket{m}^{A} \otimes \ket{\varphi_m}^{A'},
$
the \emph{quantum code} ${(\ket{\varphi_m})}_m$ being defined by
\beq
\ket{\varphi_m} =
{\frac{1}{ \sqrt S}} \sum_s \ket{\phi'_{u^{ms}}}.
\label{qcode}
\eeq
Upon applying the channel this becomes
\beq
\ket{\Psi}^{ABE} = {\frac{1}{\sqrt M}}
\sum_m \ket{m}^{A} \otimes {\frac{1}{\sqrt S}} \sum_s \ket{\phi_{u^{ms}}}^{BE}.
\label{stanje}
\eeq
The remainder of the protocol consists of Bob decoding operation.
He performs the measurement of $ms$ \emph{coherently} by appending a register
$B'$ in a standard state $\ket{0}$,  performing a unitary operation that places
the measurement ``outcome'' $ms$ into $B'$ and swapping the contents of
$B$ and $B'$. Since the original measurement 
yielded the correct value of $ms$ with probability $\approx 1$,
the coherent measurement can be constructed to output an approximation of
\beq
{\frac{1}{\sqrt{MS}}}
\sum_{ms} \ket{m}^{A} \otimes \ket{ms}^{B} \ket{\phi_{u^{ms}}}^{B' E}.
\label{prvi}
\eeq
This can be rewritten as
\begin{equation*}\begin{split}
  &{\frac{1}{\sqrt M}} \sum_m \ket{m}^{A} \otimes \ket{m}^{B}  
                              \ket{\widetilde{\phi}_{{m}}}^{B' E}, \text{ where}\\
  &\phantom{=}
   \ket{ \widetilde{\phi}_{{m}}}^{B' E}
    = {\frac{1}{\sqrt S}}\sum_s \ket{s}^{B'_1} \ket{\phi_{u^{ms}}}^{B'_2 E}.
\end{split}\end{equation*}
Observe that $\widetilde{\phi}_{{m}}^{E} \approx \theta^E$; 
by Uhlmann's theorem
their purifications are related as 
$\ket{ \widetilde{\phi}_{{m}}} \approx (V_m \otimes \1) \ket{\phi_\theta}$
for some unitary operator $V_m$. Performing the controlled unitary
$V = \sum_m \ket{m}\bra{m} \otimes V_m$ on the system $B B'$ finally
decouples Eve, yielding
$$
{\frac{1}{\sqrt M}} \sum_m \ket{m}^{A} \otimes \ket{m}^{B}
\ket{\phi_\theta}^{B' E}.
$$
The above protocol may be modified into an
\emph{entanglement transmission} one \cite{BKN}, 
in which case Alice \emph{encodes} a quantum state via
$\sum_m \ket{\varphi_m} \bra{m}$.
Hence our use of the term ``quantum code'' for $(\ket{\varphi_m})_m$.
 \paragraph{Entanglement distillation (2.).} We now turn to the construction
of a coherent version of the key distillation protocol. 
To each classical set $\CC_l$ corresponds a  quantum operator
$\Lambda_l = \sum_{ms} \ket{ms}\bra{u^{lms}}$. In lieu of a complete von Neumann
measurement on $A$, Alice performs a much less intrusive one composed of the
$\Lambda_l$ (there is also a ``failure'' outcome that happens with 
probability $\approx 0$), revealing a particular value of $l$ which is 
communicated to Bob using $n  H(X|B)^+$ bits. Their joint state becomes
\cite{remarq}
$$
\ket{\Psi}^{ABE} = {\frac{1}{\sqrt M}}
\sum_m \ket{ms}^{A} \otimes {\frac{1}{\sqrt S}} \sum_s \ket{\phi_{u^{lms}}}^{BE}.
$$
which differs from (\ref{stanje}) only in that $\ket{m}^A$ is replaced
by $\ket{ms}^A$. As before, Bob performs the coherent measurement, resulting
in the analogue of (\ref{prvi}):
$$
{\frac{1}{\sqrt{MS}}}
\sum_{ms} \ket{ms}^{A} \otimes \ket{ms}^{B} \ket{\phi_{u^{lms}}}^{B' E}.
\label{prvi2}
$$ 
To dispose of Alice's $s$-register in a coherent way further classical
communication is necessary. Alice performs a measurement in the
Fourier-transformed basis
$\ket{\hat{t}}=\frac{1}{\sqrt S}\sum_{s=1}^S e^{2\pi i st/S}\ket{s}$
($t=1,\ldots,S$),
and communicates the result $t$ to Bob using $n I(X;E)^+$ bits, 
who then applies the phase shift $\sum_{s=1}^S e^{2\pi i st/S}\ketbra{s}$
to the $s$--component of his 
$B$ register. This yields precisely 
(\ref{prvi}) and the rest of the protocol follows the entanglement generation
one above.
\par
It is possible to make a more direct connection between 
entanglement distillation and the quantum codes used for entanglement generation. 
Observe that the quantum code (\ref{qcode}) could have equally well been substituted 
by ${(\ket{\phi_{ltm}})}_m$ for any $t=1,\ldots,S$, $l = 1,\ldots, L$, where 
$$
\ket{\phi_{ltm}} = {\frac{1}{ \sqrt S}} \sum_s e^{2\pi i st/S}
\ket{\phi_{u^{lms}}}.
$$
The sets ${(\ket{\phi_{ltm}})}_{tm}$ and ${(\ket{\phi_{u^{lms}}})}_{ms}$
are (mutually unbiased) bases for the same space $\CQ_l$. The $\CQ_l$ are,
in turn, a covering of $W^{\otimes n} (\CT_A)$,
$W = \sum_x \ket{\phi_x} \bra{x}$,
since the $\CC_l$ are a  covering
of $T_X$. In other words, $W^{\otimes n} (\CT_A)$ is covered by quantum codes
in much the same way that $T_X$ is covered by key generation codes.
Our entanglement distillation protocol may be viewed as
Alice collapsing her space onto some quantum code via a measurement,
sending $I(A;E)^+$ bits of ``which code'' information to Bob,
and Bob decoding as in
entanglement generation. This view is very similar to the 
approach pursued by the Horodeckis \cite{scoop}, the difference lying in our using
``random CSS codes'' \cite{devetak} rather than random subspace ones 
\cite{shor:Q, Lloyd:Q}. 
\par
In conclusion, we have seen that all four scenarios allow an 
asymptotic conversion rate $R = I_c(A \rangle B)$, the latter
referring to the state $\rho^{AB}$ which is either given (static case),
or can be created by an application of the channel (dynamic case).
In fact, the \emph{optimal} rates for the coherent scenarios
are also given in terms of $I_c$, but applied to an appropriately blocked 
and preprocessed state/channel \cite{HHH, devetak, devetak:winter}. The same
is true for the private scenarios with $I_c(A \rangle B)$ replaced
by the possibly larger $I(X;A) - I(X;B)$ \cite{mixed}.
It remains an open question whether coherent (quantum) and private 
information are fully equivalent within our model.
\par
We thank A. Harrow, D. Leung and P. Hayden for useful comments and 
discussions.
ID is supported by the NSA under the ARO grant numbers DAAG55-98-C-0041 
and DAAD19-01-1-06. AW is supported by the U.K.~Engineering and 
Physical Sciences Research Council.


\begin{thebibliography}{M}
\bibitem{remarq} In fact, the distribution over the $m$ is
    not exactly even as in the dynamic case and the obtained
    secret key (entanglement) needs to be further ``concentrated''.
    In \cite{devetak:winter} this issue is avoided by measuring the ``type''
    of $X$ ($A$) from the start.
\bibitem{mixed} The $\phi'_x$ may be taken to be mixed if coherence
    is not required, corresponding to non-maximal measurements in
    key distillation.
\end{thebibliography}
\end{document}